\documentclass[12pt]{iopart}

\usepackage{iopams}
\usepackage{cite}
\usepackage{verbatim}
\usepackage{microtype}
\usepackage{graphicx}
\usepackage{booktabs}
\usepackage{subfigure} 
\usepackage{sidecap}
\usepackage[labelfont=bf,labelsep=period,justification=justified]{caption}

\begin{document}

\title[Nonspecific TF binding reduces variability in TF and target protein expression]{Nonspecific transcription factor binding reduces variability in transcription factor and target protein expression}

\author{M Soltani$^1$, P Bokes$^2$,
Z Fox$^3$ and A Singh$^4$}

\address{$^1$ Department of Electrical and Computer Engineering, University of Delaware, Newark, DE USA 19716.}
\address{$^2$ Department of Applied Mathematics and Statistics, Comenius University, Bratislava, Slovakia.}
\address{$^3$ Department of Chemical Biological Engineering, Colorado State University, Fort Collins, CO USA 80523.}
\address{$^4$ Department of Electrical and Computer Engineering, Biomedical Engineering, Mathematical Sciences, Center for Bioinformatics and Computational Biology, University of Delaware, Newark, DE USA 19716.}
\ead{absingh@udel.udu}
\begin{abstract}
Transcription factors (TFs) interact with a multitude of binding sites on DNA and partner proteins inside cells. We investigate how nonspecific binding/unbinding to such decoy binding sites affects the magnitude and time-scale
 of random fluctuations in TF copy numbers arising from stochastic gene expression. A stochastic model of TF gene expression, together with decoy site interactions is formulated. Distributions for the total (bound and unbound) and free (unbound) TF levels are derived by analytically solving the chemical master equation under physiologically relevant assumptions. Our results show that increasing the number of decoy binding sides considerably reduces stochasticity in free TF copy numbers. The TF autocorrelation function reveals that decoy sites can either enhance or shorten the time-scale
of TF fluctuations depending on model parameters. To understand how noise in TF abundances propagates downstream, a TF target gene is included in the model. Intriguingly, we find that noise in the expression of the target gene decreases with increasing decoy sites for linear TF-target protein dose-responses, even in regimes where decoy sites enhance TF autocorrelation times. Moreover, counterintuitive noise transmissions arise for nonlinear dose-responses. In summary, our study highlights the critical role of molecular sequestration by decoy binding sites in regulating the stochastic dynamics of TFs and target proteins at the single-cell level.
\end{abstract}
% Please keep the Author Summary between 150 and 200 words
% Use first person. PLOS ONE authors please skip this step. 
% Author Summary not valid for PLOS ONE submissions.   
%\section*{Author Summary}
\pacs{87.10.+e, 87.15.Aa, 05.10.Gg, 05.40.Ca,02.50.-r}% PACS, the Physics and Astronomy
\noindent{\it Keywords\/}: stochastic gene expression, chemical master equation, decoy binding sites, molecular sequestration, noise buffering, moment dynamics.

%Use showkeys class option if keyword
\submitto{\PB}                              %display desired
\maketitle

\section{Introduction}

Noise in the gene expression process manifests as stochastic fluctuations in protein copy numbers inside individual cells \cite{Eldar:2010kk,rao08,bkc03,keb05,rao05,mno12,arm98}. These fluctuations can be detrimental to the functioning of essential proteins whose concentrations have to be maintained within certain bounds for optimal performance \cite{lps07,fhg04,leh08}. Moreover, many diseased states have been attributed to increased noise levels in particular genes \cite{ksv02,cgt98,bhr06}. Not surprisingly, cells use a variety of regulatory mechanisms, such as incoherent feedforward circuits \cite{osella_role_2011,bleris_synthetic_2011} and negative feedback loops to minimize randomness in protein levels \cite{sak06,sih09c, lvp10, bhj03,pep08,moa04,swa04,tho01,bes00,pep08}. Here we explore an alternative noise-buffering mechanism in transcription factors (TFs): nonspecific binding of TFs to the large number of sites on DNA, referred to as \emph{decoy binding sites} \cite{wunderlich_2009}. 
%This decoy sites can be non-functional parts of DNA \cite{horng_2003}, or genes in which share the same TF \cite{Ryden14,brewster14}.

Studies have found that TF sequestration by decoy binding sites can considerably affect gene network dynamics by slowing responses times \cite{jayanthi_retroactivity_2013}, and converting graded TF-target protein dose-responses to binary responses \cite{lu_ultrasensitive_2012,chen_sequestrationbased_2012,lee2012regulatory}. Unspecific binding of TFs can also alter their stochastic dynamics. 
Using Fokker-Plank approximation to solve master equation, binding/unbinding to decoy sites was shown to reduce the magnitude of random fluctuations in TF levels \cite{burger_abduction_2010,burger_influence_2012}. Moreover, the distribution of free TF copy numbers approaches a Poisson distribution in the limit of large number of decoy sites \cite{burger_abduction_2010,burger_influence_2012}.

To understand how unspecific binding affects stochastic expression of a given TF, closed-form analytical formulas for the probability distribution, statistical moments, and the autocorrelation function of the TF population count are derived in the presence of decoy sites. Our analysis reveals that while decoy sites reduce the extent of random fluctuations, they can both shorten or lengthen the time-scale of fluctuations in the levels of the free (unbound) TF. We discuss how changes in the TF autocorrelation times by decoy sites lead to counterintuitive TF-target gene noise transmission. 

\section{Model formulation}
A schematic of the model is illustrated in Figure 1. We assume that the TF  mRNA half-life is considerably shorter than the protein half-life. In this limit, mRNAs degrade instantaneously after synthesizing a burst of protein molecules  \cite{shs08,sih09b}. TF expression is modeled as a bursty birth-death process, where TF bursts occur at a rate $k_x$ (defined as the burst frequency), and each burst generates $B_x$ molecules. Consistent with measurements \cite{YuXiao_pom_2006}, $B_x$ is assumed to be a geometrically distributed random variable with distribution
\begin{equation}
\fl {\rm Probability}\{B_x=i\}=\alpha_x(i)=(1-s_x)^{i-1}s_x, \ \ 0<s_x\leq1, \ \   i\in[1,2,\ldots).
\label{1111}
\end{equation}
%\vspace*{-25pt}
The mean burst size is given by $\langle B_x \rangle:=1/s_x \geq 1$, where $\langle . \rangle$ represents the expected value. Note that for the above burst distribution $\langle B_x \rangle=1$ if and only if
$B_x=1$ with probability one. Each TF is assumed to decay at a constant rate $\gamma_x$. Expressed TFs bind/unbind to a set of decoy binding sites with rates $k_b$ and $k_u$, respectively (Figure 1). The total number of decoy binding sites in the cell is fixed and denoted by  $N$. As in previous work, bound TFs are assumed to be protected from degradation \cite{abu_hatoum_degradation_1998,burger_abduction_2010,burger_influence_2012}. As a consequence, the average number of free TF molecules at steady-state is independent of $N$ and given by $k_x\langle B_x \rangle/\gamma_x $ \cite{burger_abduction_2010,burger_influence_2012}. 

\begin{figure*}[thp]
\center
\includegraphics[width=\textwidth]{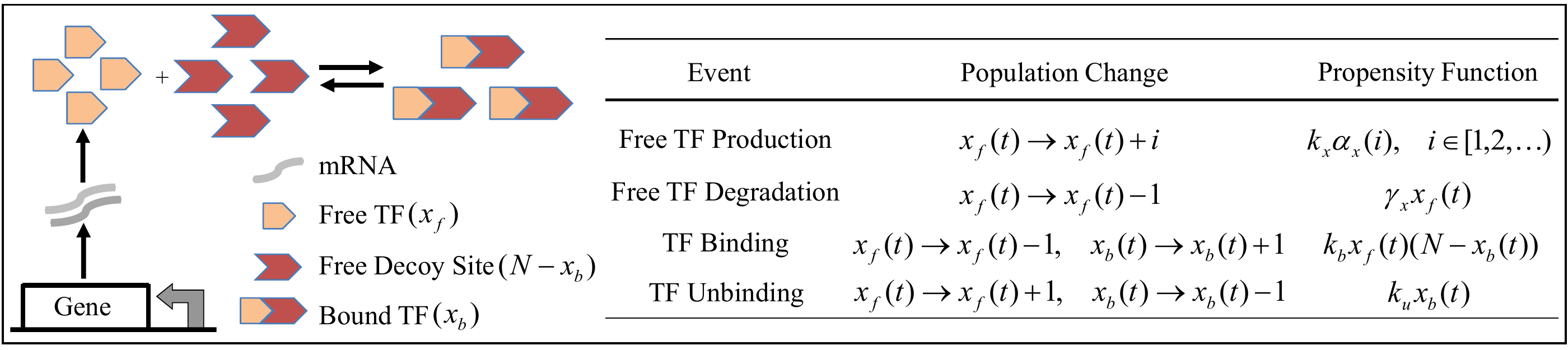}
\caption{{\bf Model schematic of transcription factor expression and interaction with decoy binding sites}. TFs are expressed from a constitutive gene, and bind/unbind to $N$ decoy binding sites with rates $k_b$ and $k_u$. The stochastic model consists of four events that ``fire" probabilistically at exponentially-distributed time intervals. Whenever an event occurs, the state of the system resets based on the second column of the table. The first event denotes protein production in bursts, with burst size distribution \eref{1111}. The cellular abundance of free and bound TF at time $t$ is represented by $x_b(t)$ and $x_f(t)$, respectively. }
\label{fig:mode}
\end{figure*}

Our model of TF expression and sequestration at decoy binding sites is based on the standard stochastic formulation of chemical kinetics \cite{mcq67,gil01}. The model is comprised of four events that occur probabilistically at exponentially-distributed time intervals (see table in Figure 1). Let $x_b(t)$, $x_f(t)$ and $x(t):=x_f(t)+x_b(t)$ denote the level of free, bound and total (free+bound) TF at time $t$ inside the cell, respectively. Then, whenever an events occurs, these population counts change based on the stoichiometry of the reaction (second column of the table). The third column lists the event propensity function $g(x_f,x_b)$, which determines how often the reactions occur. In particular, the probability that an event occurs in the next infinitesimal time interval $(t,t+dt]$ is $g(x_f,x_b)dt$. Note that the propensity function for the binding event is nonlinear and proportional to the product of $x_f$ (unbound TF) and $N-x_b$ (unbound binding sites). Our goal is to characterize the statistical properties of $x_f(t)$ when $N$ is large and is of the same order of magnitude as $\langle x_f(t) \rangle$.
A summary of notation used in the paper is provided in Table 1. Steady-state distributions of $x$ and  $x_f$ are derived next.
\begin{small}

\begin{table}[!thp]
\label{tab:parameters}
\caption{\bf{Summary of notation used in this paper.}}
%\begin{ruledtabular}
%\begin{indented}\item[]
\begin{tabular}{cl||cl }
 Parameter &Description& Parameter &Description\\
\hline
$x_f$&Free TF number&$N$&Total number of decoy sites\\
$x_b$& Bound TF number&$f$&Fraction of bound decoy sites\\
$x$&Total number of TF&$k_yx_f(t)$&Target protein burst frequency\\
$y$&Target protein number&$B_y$&Target protein burst size\\
$k_x$&TF burst frequency&$\gamma_y$&Target protein degradation rate\\
$B_x$&TF burst size&$\langle . \rangle$ & Expected value at time $t$\\
$\gamma_x$ &TF degradation rate&$CV^2$&Coefficient of variation squared\\
$k_b$&TF binding rate&$\overline{\langle . \rangle}$ &Expected value at steady-state\\
$k_u$&TF unbinding rate&$\sigma^2$ & Variance\\
 $k$ &  Dissociation constant &\\
\hline
%\end{indented}
\end{tabular}
%\end{ruledtabular}
\end{table}
\end{small}
\section{TF pdf in the presence of decoy binding sites}

If TF production is a Poisson process ($B_x=1$ with probability one), then $x_f$ has a steady-state Poisson distribution with mean $k_x/\gamma_x$, irrespective of $N$~\cite{ghaemi_stochastic_2012}. If the TF is produced in geometric bursts~\eref{1111}, an exact formula for the steady-state distribution is, as far as we know, unavailable; however, assuming that (i) the mean burst size $\langle B_x \rangle$ is large and that (ii) the TF--binding site (TF--BS) interaction rapidly equilibrates, we will show that the full bursting model, as specified by the interactions in Figure 1 and~\eref{1111}, can be approximated by a reduced model which is exactly solvable.

\subsection{Reduced model}

If $\langle B_x \rangle\gg 1$, then bursts are typically large, while decay and binding site interactions only involve one TF at a time. Thus, the contribution of bursty production to the overall gene expression noise will dominate the contributions by the decay and decoy site interactions. Because of this disparity, we treat 
the protein level as a continuous variable, which, between individual burst events, evolves deterministically in time according to rate equations which incorporate protein decay and the decoy site interactions.

Assuming that the TF--BS interaction equilibrates rapidly, the levels $x_f$ of free TF, $x_b$ of bound TF, and $N-x_b$ of free binding sites satisfy
\begin{equation}
\label{qss}
% k_b x_f(N-x_b) = k_u x_b \Rightarrow  
x_f(N-x_b) = k x_b,
\end{equation}
where $k=k_u/k_b$ is the dissociation constant.
Using $ x = x_f + x_b$ and~\eref{qss} we obtain 
\begin{equation}
x_b = \frac{N x_f }{x_f + k},\quad\Rightarrow\quad  x = x_f\left( 1 + \frac{N}{x_f + k}\right).
\label{X_from_x}
\end{equation}
The inverse relationship to~\eref{X_from_x},
\begin{equation}
 x_f = \frac{x - N - k + \sqrt{ (k + N - x)^2 + 4 k x }}{2},
 \label{x_from_X1}
\end{equation}
gives the abundance of free TF if the total TF level $x$ is given. Since binding sites protect the TF from degradation,
the rate of degradation $c=c(x)$ is proportional to the level of free TF,
\begin{eqnarray}
\label{c}
&& c(x) = \gamma_x x_f  = \frac{\gamma_x}{2}\left(x - N - k + \sqrt{ (k + N - x)^2 + 4 k x }\right).
\end{eqnarray}

The reduced model, where $x(t)$ is a continuous-state random process is given by 
\begin{equation}
\label{continuous_process}
\fl \frac{d x}{d t} = -c(x) \ \ {\rm for } \ \ t_{i-1} < t < t_{i},\quad P( x(t_i^+) > a | x(t_{i}^-) = b ) = {\rm e}^{-(a-b)/\langle B_x \rangle}, 
%&P( t_i > \tau | t_{i-1} = s) = {\rm e}^{-k_x(\tau-s)}
\end{equation}
and consists of nonlinear deterministic decay with stochastic protein bursts occurring at times $t_i, \ i=\{1,2\ldots\}$. Here $x(t_i^-)$ and $x(t_i^+)$ are the left and right limits of $x(t)$ at $t_i$. Since $x(t)$ is a continuous-state process, the geometric distribution~\eref{1111} of protein burst size has been replaced by its continuous counterpart, the exponential distribution~\cite{bokes2012multiscale, cai2006spe, friedman2006lsd}. The reduced model, belongs to a wider class of stochastic models, known as stochastic hybrid systems~\cite{sih10a}. Below we formulate and solve a master equation corresponding to this hybrid system.

\subsection{Chemical master equation with nonlinear degradation}
\label{sec:master}

The probability density function (pdf) $p(x,t)$ of observing the TF level at  $x$ at time $t$ for model \eref{continuous_process} satisfies the \emph{continuous} chemical master equation~\cite{bokes2014protein, friedman2006lsd, bokes2013transcriptional}
\begin{equation}
\label{master}
\fl \frac{\partial p(x,t)}{\partial t} -  \frac{\partial}{\partial x}(c(x)p(x,t))=  k_x \int_0^x \left( \langle B_x \rangle^{-1}e^{-\frac{(x-x')}{\langle B_x \rangle}} - \delta(x - x')\right) p(x',t) dx',
\end{equation}
subject to an initial condition
$
 p(x,t_0) = \delta(x-x_0).
$
The advective term $\partial(cp)/\partial x$ describes the transport of probability mass due to
the deterministic flow, while the integral term on the right-hand side of~\eref{master}
gives the rate of transfer of probability mass due to exponentially distributed bursts of protein synthesis~\cite{friedman2006lsd, bokes2013transcriptional}.

\begin{figure*}[h]
\center
\includegraphics[width=\textwidth]{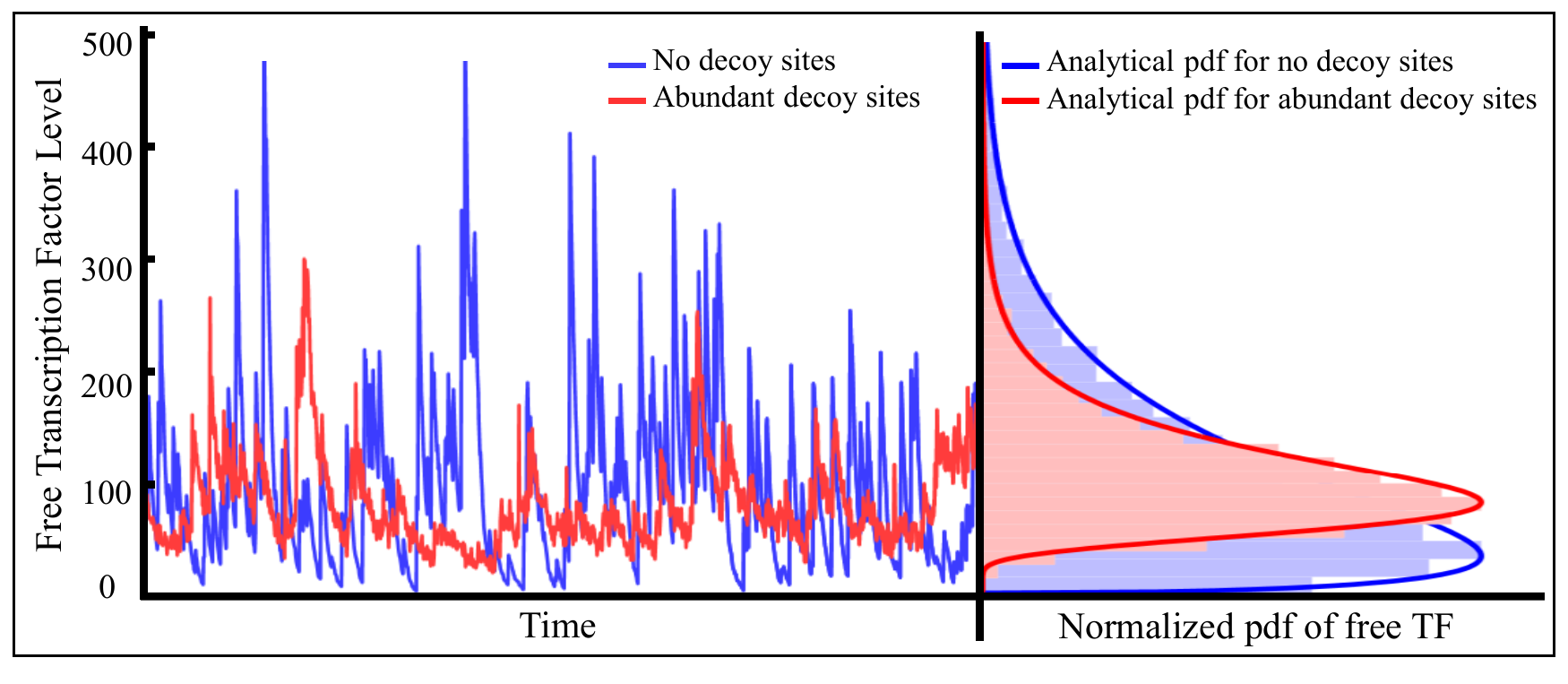}
\caption{\textbf{For large TF translational  burst sizes, adding decoy sites reduces stochasticity in free TF copy numbers}. \textit{Left:} Simulated time-evolution of free TF abundance with (red line) and without (blue line) decoy sites. Trajectories correspond to a single Monte Carlo simulation run based on the stochastic simulation algorithm\cite{Gillespie76} (SSA). \textit{Right:} Steady-state free TF distributions obtained from a large number of Monte Carlo simulation runs. Distributions obtained using the analytical formula \eref{free protein pdf} (solid lines) have an excellent match with the simulated data. Adding decoy sites considerably reduces the magnitude of fluctuations in free TF copy numbers while keeping the mean levels fixed. In this plot, the mean TF burst size  $\langle B_x \rangle =70$, and the burst frequency is adjusted so as to have on average, 
$100$ free TF molecules. Time is normalized by the mean TF life span, which is assumed to be $1/\gamma_x=1$.
 When there are no decoy sites $N=0$, and in abundance of decoy sites $N=1000$. The binding/unbinding rates were chosen such that at steady-state, the fraction of decoy sites bound with TF was $0.5$.
}
\label{fig:distributio}
\end{figure*}

When $N=0$, then $c(x)= \gamma_x x$ is linear, and the steady-state solution of \eref{master} was shown to be a gamma distribution~\cite{friedman2006lsd}. We extend this analysis to the case of nonlinear decay in \eref{c}. The distribution of the total number of TFs $x$ is (see Appendix A in SI)
\begin{equation}
\label{simplified_p}
p(x) = L x_f^{\frac{k_x}{\gamma_x} - 1} e^{\left(-\frac{x}{\langle B_x \rangle} + \frac{k_x N}{\gamma_x(x_f+k)}\right)}
	  \left(\frac{x_f}{x_f+k}\right)^{\frac{k_x N }{\gamma_x k}},
\end{equation}
where $L$ is a normalization constant, and $x_f$ is understood to be a function of $x$, as given by~\eref{x_from_X1}. The pdf $\tilde{p}(x_f)$ of observing the free TF level at $x_f$ is obtained from~\eref{simplified_p} using the transformation rule
\begin{equation}
\eqalign{
\fl \tilde{p}(x_f) = p(x(x_f))\frac{d x}{d x_f} =\\
 L x_f^{\frac{k_x}{\gamma_x} - 1} \rme^{\left(-\frac{x_f}{\langle B_x \rangle}-\frac{Nx_f}{\langle B_x \rangle \left(x_f + k\right)} + \frac{k_x N}{\gamma_x(x_f+k)}\right)}
	  \left(\frac{x_f}{x_f+k}\right)^{\frac{k_x N }{\gamma_x k}}\left(1 + \frac{N k}{(x_f + k)^2}\right), 
%  p(x(x_f))\left(1 + \frac{N k}{(x_f + k)^2}\right)
   \label{free protein pdf}
   }
\end{equation}
where $p(x(x_f))$ is given by~\eref{simplified_p}, wherein $x_f$ becomes the independent
variable, while $x$ is understood to be a function of $x_f$, as given by~\eref{X_from_x}. The distribution of free TF based on the above formula matches very well with distributions obtained from running a large number of Monte Carlo simulations (Figure \ref{fig:distributio}). Our results show that increasing the number of decoy sites considerably reduces stochastic variability in the free TF population counts (Figure \ref{fig:distributio}). 

\section{Noise level of TF and target protein }
Next we investigate how noise in TF levels propagates downstream to target proteins. To do so we consider a target protein activated by the free TF via a linear dose-response.
The stochastic model for target protein activation is given by 
\begin{equation}
\eqalign{
{\rm Probability}\{y(t+dt)=y(t)+i \}=k_y x_f(t) \alpha_y(i) dt,\\
{\rm Probability}\{y(t+dt)=y(t)-1 \}=\gamma_y y(t) dt,
} \label{chemical reactions}
\end{equation}
where $x_f(t)$ and $y(t)$ denote the free TF level and the target protein level at time $t$, respectively, and $\gamma_y$ is the degradation rate. Target protein is expressed in bursts, with burst frequency $k_y x_f(t)$, and each burst generates $B_y$ geometrically distributed molecules
\begin{equation}
\fl {\rm Probability}\{B_y=i\}=\alpha_y(i)=(1-s_y)^{i-1}s_y, \ \ 0<s_y\leq1, \ \   i\in[1,2,\ldots).
\label{1112}
\end{equation}
The overall system consists of the table in Figure \ref{fig:mode} and equation \eref{chemical reactions}. To quantify $y(t)$ noise level, time evolution of un-centered statistical moments of the stochastic processes $x_f(t)$, $x_b(t)$ and $y(t)$ are first derived. Moment dynamics is obtained using the following result:
based on Theorem 1 of \cite{RNC:RNC1017} the time derivative of the expected value of any function $\varphi(x_f,x_b,y)$ is given by 
\begin{equation}
\frac{d\langle \varphi(x_f,x_b,y) \rangle}{dt}= \left \langle \sum_{Events}  \Delta \varphi(x_f,x_b,y) \times g(x_f,x_b,y) \right \rangle,   
\label{dynnf}
\end{equation}
where $\Delta \varphi(x_f,x_b,y)$ is a change in $\varphi$ when an event occurs, and $g(x_f,x_b,y)$ is the event propensity function \cite{RNC:RNC1017}. However, because of the nonlinear propensity function of the TF binding event, $k_b x_f(N-x_b)$, we encounter the well-known problem of moment-closure: the time derivative of lower-order moments depend on higher-order moments \cite{sih10a,sih10}. In such cases moments are typically obtained using different moment closure schemes \cite{gov06,lkk09,gou05,sih10,gillespie2009}. Here we use the well-known linear noise approximation (LNA), where the mean population counts are identical to the deterministic chemical rate equations \cite{Kampen01}. Based on this method, we linearize this propensity function around the steady-state mean levels, i.e.,
 \begin{equation}
k_b x_f(N-x_b)\approx k_b N x_f- k_b \left(   x_f \overline{ \langle x_b \rangle} + \overline{\langle x_f \rangle} x_b - \overline{\langle x_f \rangle}\hspace{1mm}\overline{\langle x_b \rangle}\right),
\label{lineraization}
\end{equation}
where $ \overline{\langle x_f \rangle}$ and $\overline{\langle x_b \rangle}$ denote the steady-state mean levels of the free and bound TF, respectively. 

Using \eref{lineraization} in place of the original nonlinear propensity function, closed moment dynamics is obtained by appropriately choosing $\varphi(x_f,x_b,y)$ in \eref{dynnf} (see Appendix B in SI). Steady-state analysis of moment dynamics results in the following noise levels measured by the steady-state coefficient of variation ($CV$) squared (variance/mean$^2$) for the free TF, bound TF and target protein:
\numparts
\begin{eqnarray}
 CV^2_{x_f} = \frac{1}{\overline{\langle x_f \rangle}}+ \frac{\langle B_x \rangle-1}{Nf(1-f)+\overline{\langle x_f \rangle}},\label{cvxf}\\
 CV^2_{x_b} = \frac{(1-f)\left(  \overline{\langle x_f \rangle}+ Nf(1-f)\langle B_x \rangle \right) }{ N f(Nf(1-f)+ \overline{\langle x_f \rangle})},\label{cvxb}\\
 CV^2_y =  \frac{\langle B_y\rangle}{\overline{\langle y\rangle}}+  \frac{\langle B_x \rangle \gamma_y}{\gamma_y (f (1-f) N + \overline{\langle x_f \rangle})+\gamma_x \overline{\langle x_f \rangle}},\label{cvy}
\end{eqnarray}
\endnumparts
where 
$
f=\overline{\langle x_f \rangle}/( k+\overline{\langle x_f \rangle})
$
is the fraction of decoy sites that are occupied at steady state (Appendix B in SI). The steady-state means are given by
\begin{equation}
\overline{\langle x_f \rangle }= \frac{k_x \langle B_x \rangle}{\gamma_x}, \ \, \overline{\langle x_b \rangle}=Nf, \ \ \overline{\langle y \rangle} = \frac{k_y \overline{\langle x_f \rangle } \langle B_y \rangle}{\gamma_y}.
\label{mean1}
\end{equation}
In addition, the covariance between free and bound TF is computed as 
\begin{equation}
Cov(x_f,x_b)= \frac{Nf(1-f)\overline{\langle x_f \rangle}(\langle B_x \rangle-1)}{Nf(1-f)+\overline{\langle x_f \rangle}}.
\label{crosscorrelation}
\end{equation}
\begin{figure}[!h]
\center
\includegraphics[width=\textwidth]{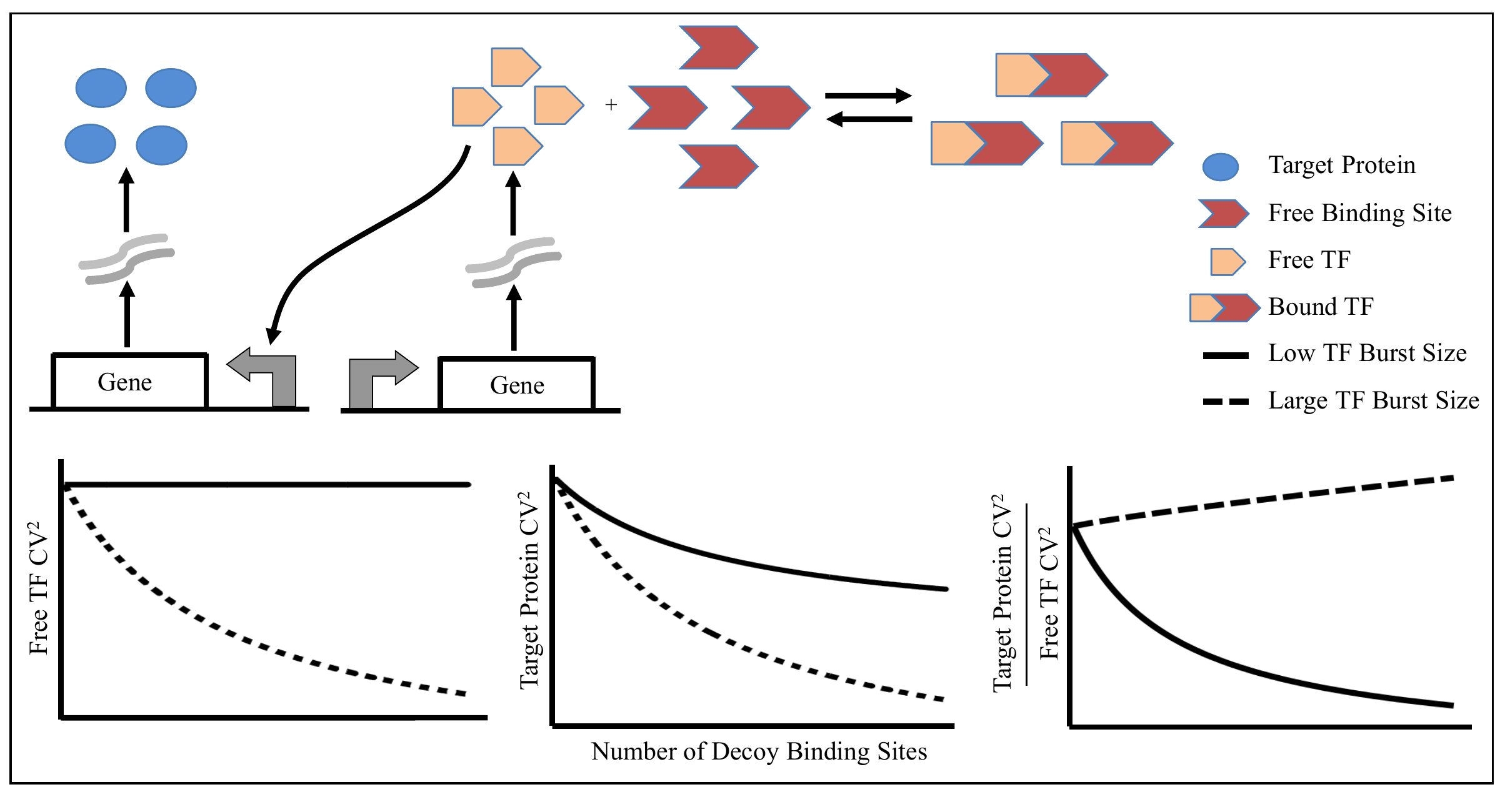}
\caption{\textbf{Addition of decoy binding sites reduces noise in target protein expression.} \textit{Top:} Model schematic of TF activating a target protein. Target protein production rate is assumed to be linearly dependent on  the free TF abundance. \textit{Bottom:} Qualitative plots of free TF $CV^2$ (coefficient of variation squared), target protein $CV^2$, and their ratio as a function of  the number of TF decoy binding sites $N$. For low TF burst sizes (solid line; $B_x=1$ with probability one), noise in free TF levels is invariant of  $N$, while target protein $CV^2$ monotonically decreases with $N$. For large burst sizes (dashed line; $\langle B_x \rangle > 1$), both free TF and target protein $CV^2$ reduce with increasing $N$. The ratio of free TF and target protein $CV^2$ decreases (increases) with $N$ for low (high) burst size. Noise levels are normalized by their values when there are no decoy sites ($N=0$). Parameters chosen as in Figure 2. Target protein mean is 50 and its degradation rate was assumed to be similar to that of transcription factor.}
\label{fig:cvxovercvz}
\end{figure}
The following observations can be made from (14)-\eref{crosscorrelation}:
\begin{itemize}
\item When $\langle B_x \rangle=1$, free TF has Posisson statistics, and noise level $CV^2_{x_f}=1/{\overline{\langle x_f \rangle}}$ is independent of  $N$. In contrast, noise in the target protein decreases with increasing $N$ (Figure 3).
\item For large burst sizes ($\langle B_x \rangle>1$), noise in both free TF and target protein populations decreases with increasing number of decoy sites (Figure 3). 
\item Because of the term $f(1-f)$, decrease in the noises of both free TF and target protein are maximal when $f=0.5$ (half of the total sites are occupied).
\item For $\langle B_x \rangle>1$ and $0<f<1$
\begin{equation}
\lim_{N \to \infty} CV^2_{x_f} \to \frac{1}{{\overline{\langle x_f \rangle}}}, \ \  \lim_{N \to \infty} CV^2_{y} \to 0.
\end{equation}
\item The ratio $CV^2_y/CV^2_{x_f}$ decreases (increases) with $N$ for small (large) TF burst size (Figure 3). 
\item When $\langle B_x \rangle=1$, $Cov(x_f,x_b)=0$, i.e., bound and free TF levels are uncorrelated at steady-state. The correlation between $x_f$ and $x_b$ monotonically increases with mean TF burst size (Figure 4).
\end{itemize}

To understand some of these results, such as why variability in target protein expression attenuates with increasing $N$ for $\langle B_x \rangle=1$, while noise in the free TF population remains fixed, we investigate 
the $x_f(t)$ autocorrelation function.

\begin{SCfigure}[][h]
\includegraphics[width=8.9cm,clip]{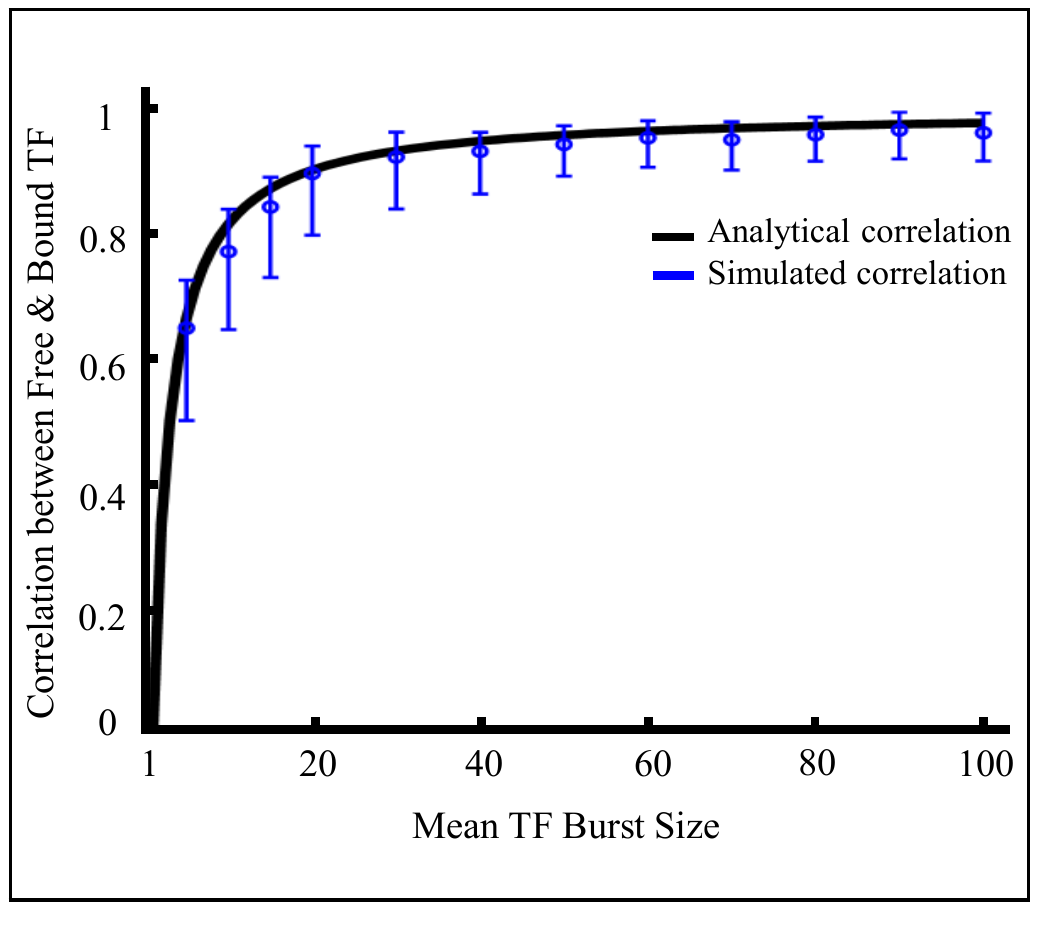} 
\caption{\textbf{Correlation between free and bound TF increases with mean TF burst size.} When the TF burst size is $B_x=1$ with probability one (i.e., TF production is a Poisson process), bound and free TF levels are uncorrelated.  As the mean TF burst size increases, the correlation between them approaches one. Parameters are chosen as  $\overline{\langle x_f \rangle }=100$, $f=0.5$ and $N=500$. The results from $1000$ Monte Carlo simulations are also shown with a 95\% confidence interval calculated using bootstrapping.}
\label{fig:correlation}
\end{SCfigure}

\section{Free TF Autocorrelation function}

The steady-state autocorrelation function for the free TF abundance $x_f(t)$ is defined as
\begin{equation}\label{auto corr0}
R(t):=\frac{\langle x_f(t+s)x_f(s) \rangle - \overline{\langle x_f \rangle}^2}{\overline{\langle x_f^2 \rangle} - \overline{\langle x_f \rangle}^2}.
\end{equation}
In the case of no decoy sites ($N=0$), the autocorrelation function is

%\subsection*{Absence of decoy binding sites}
%
%To compute the auto-correlation function we use the fact that
%\begin{equation}
%        \begin{aligned}
%& \langle x_f(t+s)x_f(s) \rangle =\langle x_f(s) \langle  x_f(t+s) \vert x_f(s)  \rangle \rangle,
%  \label{ts0}
%        \end{aligned}
%\end{equation}
%where $\langle  x_f(t+s) \vert x_f(s)  \rangle$ is the expected number of free TF molecules at time $t+s$ given $x_f(s)$ \cite{singh_consequences_2012}.
%When there are no decoy sites, the mean TF population count evolves according to the following linear differential equation 
%\begin{align}
%&  \frac{d\langle x_f \rangle}{dt}=k_x \langle B \rangle -\gamma_x \langle x_f\rangle,
%\end{align}
%which implies
%\begin{equation}\label{cdm}
%        \begin{aligned}
%&\langle x_f(t+s) \vert x_f(s)  \rangle= \overline{\langle x_f \rangle} + \left( x_f(s)- \overline{\langle x_f \rangle} \right) e^{-\gamma_x  t}. \\
%        \end{aligned}
%\end{equation}
%Using \eref{ts0} and \eref{cdm}
%\begin{equation}
%        \begin{aligned}
%&\langle x_f(t+s)x_f(s) \rangle = \overline{\langle x_f \rangle}^2  + \left( \overline{\langle x_f^2 \rangle} - \overline{\langle x_f \rangle}^2  \right) e^{-\gamma_x t}.
%  \label{ts2}
%        \end{aligned}
%\end{equation}
%Substituting \eref{ts2} in \eref{auto corr0} we obtain the autocorrelation function 
\begin{equation}\label{auto corr00}
 R(t)=\rme^{-\gamma_x t},
\end{equation} 
which is completely determined by the TF decay rate \cite{singh_consequences_2012}.

In the presence of binding sites, the system has two time-scales: fast binding/unbinding of TF to decoy sites, and slow TF production/degradation. Given $x_f(s)$ and $x_b(s)$ at some initial time $s$, population counts
change rapidly and reach manifold \eref{X_from_x} determined by the quasi steady-state equilibrium of binding/unbinding reactions. Let $s^+$ denote a time immediately after time $s$ such that $\forall t \geq s^+$, $x_f(t)$ and $x_b(t)$ remain on the manifold, i.e., 
\begin{equation}
x(t)= x_f(t)\left( 1 + \frac{N}{x_f(t)+ k}\right).
\label{s to splus}
\end{equation}
Moreover, the total TF abundance $x(s)=x(s^+)$ because there are no TF birth/death events in this short time. After the initial fast change, autocorrelation function is defined by 
\begin{equation}
\label{auto corr}
\fl R(t)=\frac{\langle x_f(s^+)x_f(t+s^+) \rangle - \overline{\langle x_f \rangle}^2}{\overline{\langle x_f^2 \rangle} - \overline{\langle x_f \rangle}^2}=\frac{\langle x_f(s^+) \langle  x_f(t+s^+) \vert x_f(s^+)  \rangle \rangle - \overline{\langle x_f \rangle}^2}{\overline{\langle x_f^2 \rangle} - \overline{\langle x_f \rangle}^2}.
\end{equation}
We use conditioning to express the term $\langle x_f(s^+)x_f(t+s^+) \rangle$ based on $x_f(s^+)$. To obtain the conditional mean $ \langle  x_f(t+s^+) \vert x_f(s^+)  \rangle$, we derive the time evolution of $x_f(t)$ on the manifold \eref{s to splus} (see Appendix C in SI)
\begin{equation}
\frac{d \langle x_f \rangle }{d t}= \tilde{k}_x - \tilde{\gamma}_x \langle x_f \rangle 
\label{new ode}
\end{equation}
where
\begin{equation}
\tilde{k}_x:=\frac{k_x \langle B \rangle}{1 + \frac{Nf(1-f)} {\overline{\langle x_f \rangle}}}, \ \ \tilde{\gamma}_x := \frac{\gamma_x}{ 1+\frac{N f(1-f)}{\overline{\langle x_f \rangle}}},
\label{new ode1}
\end{equation}
and shows a slower convergence rate $\tilde{\gamma}_x$ of $x_f(t)$ to its steady-state compared to $\gamma_x$. From \eref{new ode},
\begin{equation}
 \langle x_f(t+s^+) \vert x_f(s^+)\rangle = \overline{\langle x_f \rangle} + \left(  x_f(s^+)  - \overline{\langle x_f \rangle}\right) e^{-\tilde{\gamma}_x t},
  \label{tsplus}
\end{equation}
which using \eref{auto corr} yields
\begin{equation}\label{auto corr2}
 R(t)=\frac{\langle x_f^2(s^+) \rangle - \overline{\langle x_f \rangle}^2}{\overline{\langle x_f^2 \rangle} - \overline{\langle x_f \rangle}^2} e^{-\tilde{\gamma}_x t}.
\end{equation}
Assuming that TF noise levels are sufficiently small, $\langle x_f(s^+)^2 \rangle$  can be approximated via a Taylor series as (see Appendix C in SI)
\begin{equation}
\langle x_f^2(s^+) \rangle =  \overline{\langle x_f \rangle}^2+\left( \frac{\overline{\langle x_f \rangle}}{N f(1-f)+\overline{\langle x_f \rangle}} \right)^2 \sigma^2_x,
\label{approximation of xf2}
\end{equation}
where $\sigma^2_x$ is the steady-state variance of the total TF abundance.
Combining equations \eref{auto corr2} and \eref{approximation of xf2}, the autocorrelation function is given by
\begin{equation}
\label{auto corr3}
 R(t)=\left( \frac{\overline{\langle x_f \rangle}}{N f(1-f)+\overline{\langle x_f \rangle}} \right)^2  \frac{\sigma^2_x}{\sigma^2_{x_f}}  e^{-\tilde{\gamma}_x  t}, 
\end{equation}
for $t>0$ and $R(0)=1$. The ratio of variances $\sigma^2_x/\sigma^2_{x_f}$ can be obtained from the mean and noise levels in (14)-\eref{crosscorrelation}. As expected, when $N=0$, \eref{auto corr3}
reduces to \eref{auto corr00}.

Systematic analysis of \eref{auto corr3} reveals that nonspecific binding either increase or decrease $\tau_{50}$ (time at which $R(t)$ reached $50\%$ of its maximum value) depending on $\langle B_x \rangle$ (Figure 5). In particular, for low TF burst size ($\langle B_x \rangle \approx 1)$, adding decoy sites makes the autocorrelation function biphasic, with a sharp initial drop followed by a slow exponential decay $ e^{-\tilde{\gamma}_x t}$. In this case, increasing $N$ shifts $\tau_{50}$ to the left (Figure 5 left). As time-scale of $x_f(t)$ fluctuations become faster with increasing $N$, variability in target protein expression decreases due to efficient time averaging of upstream TF fluctuations (Figure 3). Keeping $N$ fixed, as one increases $\langle B_x \rangle$ the initial drop reduces and the autocorrelation function becomes dominated by $e^{-\tilde{\gamma}_x t}$ (Figure 5 right). Hence, for large TF burst sizes, nonspecific TF binding can enhance $\tau_{50}$ making $x_f(t)$ fluctuations longer and more permanent.

%
%Therefore, at time zero, autocorrelation is one, and for any time ($t>0$), autocorrelation can be calculated by equation \eref{auto corr3}. From this equation, for a constant mean of free TF and fraction of bound decoy sites, two parameters ($N$ and $\langle B \rangle$) determine the behavior of the autocorrelation function.
%
%TF fluctuations have both a slow and a fast manifold. When burst size is small the fast manifold is dominant and causes fast reduction of autocorrelation at time zero, but when burst size is large, the slow manifold is dominant resulting in slow reduction of autocorrelation. In the limit of a very large burst size the autocorrelation will only move on the slow manifold, thus autocorrelation is 
%\begin{equation}
%\begin{aligned}\label{auto corr4}
%& R(t)= e^{-\tilde{\gamma}_x  t}. 
%\end{aligned}
%\end{equation}
%
%When the number of decoy sites goes to infinity, the autocorrelation trend will again move in the fast manifold.
%
%Autocorrelation is compared for low and high burst size in Figure \ref{fig:autocorrelation}. As illustrated in this figure, a change in burst size changes 

\begin{figure}[thpb]
\center
\includegraphics[width=\textwidth]{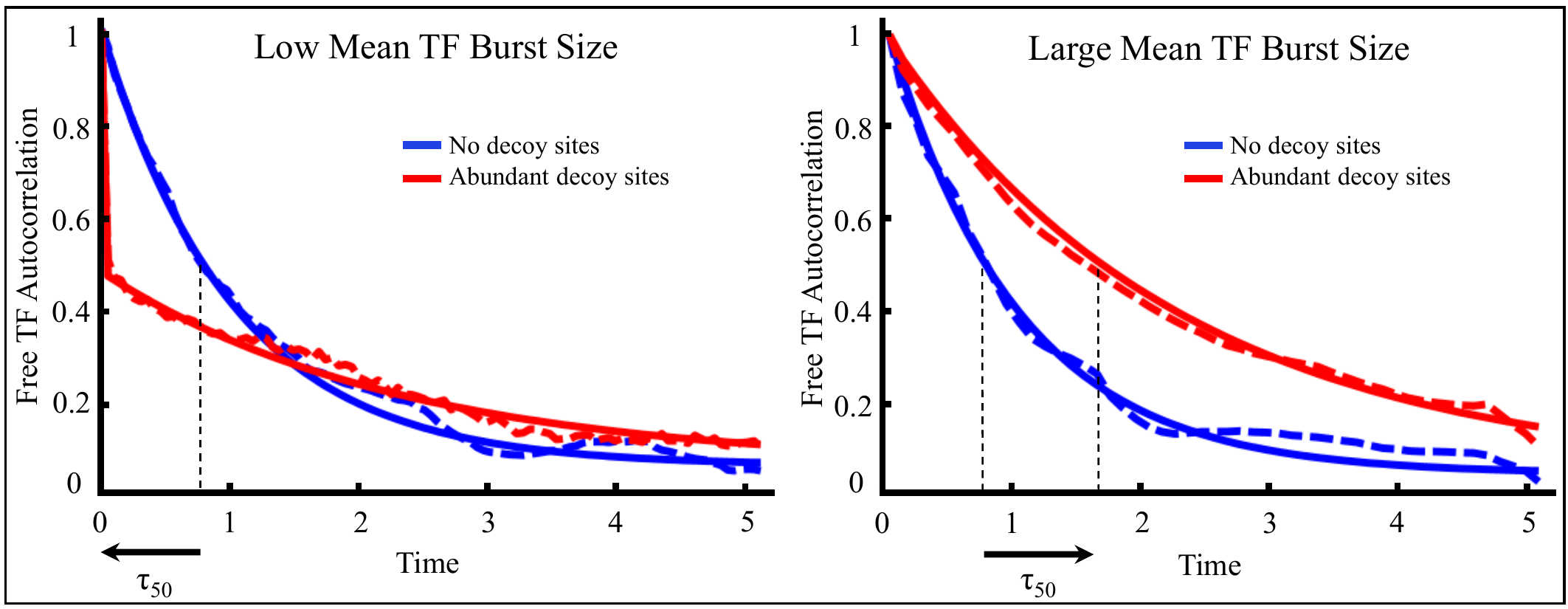}
\caption{\textbf{Decoy binding sites enhance or shorten TF autocorrelation times depending on the TF burst size.} Autocorrelation function $R(t)$ of the free TF population count for low ($\langle B_x \rangle=1$; left) and high ($\langle B_x \rangle=100$; right) TF burst sizes. For low burst sizes, adding decoy sites makes the autocorrelation function biphasic and shifts $\tau_{50}$ (time at which $R(t)=0.5$) to the left. In contrast, for large burst sizes, adding decoy sites shifts $\tau_{50}$ to the right. Solid lines represent $R(t)$ obtained from \eref{auto corr3}, while dashed lines correspond to Monte Carlo simulations. For this plot, parameters taken as $f=0.5$, $\gamma_x=1$ and $\overline{\langle x_f \rangle }=100$. For no decoy sites $N=0$, while for abundant decoy sites $N=500$.}
\label{fig:autocorrelation}
\end{figure}

\section{Discussion}

We investigated how nonspecific binding affects random fluctuations in the abundance of a given TF inside single-cells. A stochastic model of TF expression and interaction with $N$ decoy binding sites
was formulated and analyzed assuming: 1) Each gene expression event creates a geometrically distributed burst of TFs; 2) TF binding/unbinding to decoy sites is fast compared to TF production/degradation; and 3) Bound TFs are protected from degradation. The latter assumption ensures that the mean level of the free (unbound) TF is independent of $N$ at equilibrium (see \eref{mean1}).

\subsection{Effect of decoy sites on TF noise level}

Equation (14) shows that the free TF noise level (as measured by the steady-state coefficient of variation squared $CV^2_{x_f}$) is invariant of $N$ if:
\begin{itemize}
\item TF production is a Poisson process ($\langle B_x \rangle =1$).
\item Weak TF interaction with decoy sites ($f=0$; all sites are unbound).
\item Strong TF interaction with decoy sites ($f=1$; all sites are bound).
\end{itemize}
However, for $\langle B_x \rangle >1$ and $0<f<1$, $CV^2_{x_f}$ decreases with increasing number of decoy sites. Intuitively, noise reduction occurs because if there is a large expression burst by random chance, then many TFs would rapidly bind to unbound decoy sites, minimizing the magnitude of fluctuation in the free TF population. In the limit $N \to \infty$, $CV^2_{x_f}$ approaches the Poisson limit ($CV^2=1/Mean$). Note that this noise buffering comes at the cost of slower response times: After gene induction, it takes a much longer time for the amount of free TF to reach a critical threshold in the presence of decoy sites than in their absence.

Our model only takes into consideration intrinsic noise in TF synthesis, i.e., noise arising from the inherent stochastic nature of gene expression. Additional variability, referred to as \emph{extrinsic noise} \cite{sos08,hip11,ses02}, arises from fluctuations in the environment or abundance of gene expression machinery. Can nonspecific binding also reduce extrinsic noise in TF expression? We incorporated extrinsic noise in the model by assuming that the burst frequency $k_x$ is itself a stochastic process \cite{moh01}.  Monte Carlo simulations confirm that $CV^2_{x_f}$ decrease with increasing $N$ irrespective of whether gene expression noise is intrinsic or extrinsic (see Appendix D in SI). 

\subsection{TF-target protein noise transmission}

To quantify TF noise propagation downstream one needs to characterize both the magnitude and time-scale of TF copy number fluctuations. Intriguingly, we find that nonspecific TF binding can shift 
$x_f(t)$ fluctuations to slower or faster time scales depending on $\langle B_x \rangle$. When $\langle B_x \rangle \approx 1$ and $N$ is large, $x_f(t)$ autocorrelation function $R(t)$ has a rapid initial drop (Figure 5). Recall that when $\langle B_x \rangle \approx 1$, bound and free TF levels are uncorrelated (Figure 4). Thus the initial drop represents loss of temporal correlations due to rapid equilibration of binding/unbinding reactions. This initial phase is followed by an exponential decay $e^{-\tilde{\gamma}_x  t}$, which corresponds to slow convergence of $x_f(t)$ fluctuations on the manifold \eref{X_from_x}. Since for low TF burst sizes increasing $N$ shifts $x_f(t)$ to faster time-scales without altering $CV^2_f$, noise in target protein levels decreases due to efficient time averaging irrespective of the TF-target protein dose-response.

A contrasting scenario emerges when the TF burst size is large ($\langle B_x \rangle \gg 1$). In this case bound and free TF levels are highly correlated and are close to the manifold \eref{X_from_x}. Hence, the initial drop in $R(t)$ is reduced and the autocorrelation function is dominated by the slow exponential decay $e^{-\tilde{\gamma}_x  t}$ (Figure 5). For large TF burst sizes, increasing $N$ makes $x_f(t)$ fluctuations smaller (which decreases noise in target protein) and slower (which increases noise in target protein). Our analysis shows that the net effect is to reduce variability in target protein expression for linear dose-responses (Figure 3). Note that the ratio of target protein and TF noise levels increases with $N$ (Figure 3), i.e., if one were to 
increase $N$ by keeping $CV^2_f$ fixed (for example, by  simultaneously changing the TF burst size), then noise in target protein levels would increase due to less efficient time averaging of upstream TF fluctuations. Interestingly, we find that for large TF burst sizes counterintuitive noise transmissions arise when the dose-response curve is nonlinear. For example, consider a Hill function dose-response, i.e., target protein burst frequency is $k_y x_f(t)^h/(x_f(t)^h+\overline{\langle x_f \rangle }^h)$. Monte Carlo simulations reveal that when $h$ is large, increasing $N$ enhances noise in the target protein population due to longer (but smaller) fluctuations in $x_f(t)$ (see Appendix E in SI).

\section{Conclusion}

In summary our results show that nonspecific TF binding to the large number of sites on DNA plays a critical role in regulating TF copy number fluctuations inside individual cells. Moreover, noise attenuation is also achieved for target proteins as long as the TF-target gene dose-response is linear. For nonlinear dose-responses, nonspecific TF binding can amplify variability in the target protein population, even though noise in the free TF population is attenuated. Future efforts will focus on experimentally verifying these result using synthetic genetic circuits and understanding how nonspecific binding affects the stochastic dynamics of complex gene regulatory networks.

\section*{Acknowledgments}
PB was supported by the Slovak Research and Development Agency (contract no. APVV-0134-10) and also by the VEGA grant agency (contract no. 1/0711/12).
AS is supported by the National Science Foundation Grant DMS-1312926, University of Delaware Research Foundation (UDRF) and Oak Ridge Associated Universities (ORAU).

% You may title this section "Methods" or "Models". 
% "Models" is not a valid title for PLoS ONE authors. However, PLoS ONE
% authors may use "Analysis" 
%\section*{Materials and Methods}

%\section*{Figure Legends}

% This section is for figure legends only, do not include
% graphics in your manuscript file.
%
%\begin{figure}
%\caption{
%{\bf Bold the first sentence.}  Rest of figure caption.  
%}
%\label{Figure_label}
%\end{figure}

%\newpage
%\section*{Tables}
% 
% See introductory notes if you wish to include sideways tables.
%
% NOTE: Please look over our table guidelines at http://www.plosone.org/static/figureGuidelines#tables to make sure that your tables meet our requirements. Certain types of spacing, cell merging, and other formatting tricks may have unintended results and will be returned for revision.
%
%\begin{table}[!ht]
%\caption{
%\bf{Table title}}
%\begin{tabular}{|c|c|c|}
%table information
%\end{tabular}
%\begin{flushleft}Table caption
%\end{flushleft}
%\label{tab:label}
% \end{table}

%\section*{Supporting Information Legends}
%
% Please enter your Supporting Information captions below in the following format:
%\item{\bf Figure SX. Enter mandatory title here.} Enter optional descriptive information here.
% 
%\begin{description}
%\item {\bf}
%\item {\bf}
%\end{description}

\section*{References}
%\nocite{*}
\bibliographystyle{mohammad}
\bibliography{thesisJCP,thesis}% Produces the bibliography via BibTeX.

\providecommand{\newblock}{}
\begin{thebibliography}{10}
\expandafter\ifx\csname url\endcsname\relax
  \def\url#1{{\tt #1}}\fi
\expandafter\ifx\csname urlprefix\endcsname\relax\def\urlprefix{URL }\fi
\providecommand{\eprint}[2][]{\url{#2}}

\bibitem{Eldar:2010kk}
Eldar A and Elowitz M~B 2010 {\em Nature\/} {\bf 467} 167--173

\bibitem{rao08}
Raj A and van Oudenaarden A 2008 {\em Cell\/} {\bf 135} 216--226

\bibitem{bkc03}
Blake W~J, Kaern M, Cantor C~R and Collins J~J 2003 {\em Nature\/} {\bf 422}
  633--637

\bibitem{keb05}
Kaern M, Elston T~C, Blake W~J and Collins J~J 2005 {\em Nature Reviews
  Genetics\/} {\bf 6} 451--464

\bibitem{rao05}
Raser J~M and O'Shea E~K 2005 {\em Science\/} {\bf 309} 2010--2013

\bibitem{mno12}
Munsky B, Neuert G and van Oudenaarden A 2012 {\em Science\/} {\bf 336}
  183--187

\bibitem{arm98}
Arkin A, Ross J and McAdams H~H 1998 {\em Genetics\/} {\bf 149} 1633--1648

\bibitem{lps07}
Libby E, Perkins T~J and Swain P~S 2007 {\em Proceedings of the National
  Academy of Sciences\/} {\bf 104} 7151--7156

\bibitem{fhg04}
Fraser H~B, Hirsh A~E, Giaever G, Kumm J and Eisen M~B 2004 {\em PLoS
  Biology\/} {\bf 2} e137

\bibitem{leh08}
Lehner B 2008 {\em Molecular Systems Biology\/} {\bf 4} 170

\bibitem{ksv02}
Kemkemer R, Schrank S, Vogel W, Gruler H and Kaufmann D 2002 {\em Proceedings
  of the National Academy of Sciences\/} {\bf 99} 13783--13788

\bibitem{cgt98}
Cook D~L, Gerber A~N and Tapscott S~J 1998 {\em Proceedings of the National
  Academy of Sciences\/} {\bf 95} 15641--15646

\bibitem{bhr06}
Bahar R, Hartmann C~H, Rodriguez K~A, Denny A~D, Busuttil R~A, Dolle M~E,
  Calder R~B, Chisholm G~B, Pollock B~H, Klein C~A and Vijg J 2006 {\em
  Nature\/} {\bf 441} 1011--1014

\bibitem{osella_role_2011}
Osella M, Bosia C, Corá D and Caselle M 2011 {\em {PLoS} Comput Biol\/} {\bf
  7} e1001101

\bibitem{bleris_synthetic_2011}
Bleris L, Xie Z, Glass D, Adadey A, Sontag E and Benenson Y 2011 {\em Molecular
  Systems Biology\/} {\bf 7} 519

\bibitem{sak06}
El-Samad H and Khammash M 2006 {\em Biophysical Journal\/} {\bf 90} 3749--3761

\bibitem{sih09c}
Singh A and Hespanha J~P 2009 {\em IET Systems Biology\/} {\bf 3} 368--378

\bibitem{lvp10}
Lestas I, Vinnicombegv G and Paulsson J 2010 {\em Nature\/} {\bf 467} 174--178

\bibitem{bhj03}
Bundschuh R, Hayot F and Jayaprakash C 2003 {\em J. of Theoretical Biology\/}
  {\bf 220} 261--269

\bibitem{pep08}
Pedraza J~M and Paulsson J 2008 {\em Science\/} {\bf 319} 339--343

\bibitem{moa04}
Morishita Y and Aihara K 2004 {\em J. of Theoretical Biology\/} {\bf 228}
  315--325

\bibitem{swa04}
Swain P~S 2004 {\em J. Molecular Biology\/} {\bf 344} 956--976

\bibitem{tho01}
Thattai M and van Oudenaarden A 2001 {\em Proceedings of the National Academy
  of Sciences\/} {\bf 98} 8614--8619

\bibitem{bes00}
Becskei A and Serrano L 2000 {\em Nature\/} {\bf 405} 590--593

\bibitem{wunderlich_2009}
Wunderlich Z and Mirny L~A 2009 {\em Trends in genetics: {TIG}\/} {\bf 25}
  434--440

\bibitem{jayanthi_retroactivity_2013}
Jayanthi S, Nilgiriwala K~S and Del~Vecchio D 2013 {\em {ACS} Synthetic
  Biology\/} {\bf 2} 431--441

\bibitem{lu_ultrasensitive_2012}
Lu M~S, Mauser J~F and Prehoda K~E 2012 {\em {ACS} synthetic biology\/} {\bf 1}
  65--72

\bibitem{chen_sequestrationbased_2012}
Chen D and Arkin A~P 2012 {\em Molecular Systems Biology\/} {\bf 8} 620

\bibitem{lee2012regulatory}
Lee T and Maheshri N 2012 {\em Molecular systems biology\/} {\bf 8} 576

\bibitem{burger_abduction_2010}
Burger A, Walczak A~M and Wolynes P~G 2010 {\em Proceedings of the National
  Academy of Sciences\/} {\bf 107} 4016--4021

\bibitem{burger_influence_2012}
Burger A, Walczak A~M and Wolynes P~G 2012 {\em Phys. Rev. E\/} {\bf 86} 041920

\bibitem{shs08}
Shahrezaei V and Swain P~S 2008 {\em Proceedings of the National Academy of
  Sciences\/} {\bf 105} 17256--17261

\bibitem{sih09b}
Singh A and Hespanha J~P 2009 {\em Biophysical Journal\/} {\bf 96} 4013--4023

\bibitem{YuXiao_pom_2006}
Yu J, Xiao J, Ren X, Lao K and Xie X~S 2006 {\em Science\/} {\bf 311}
  1600--1603

\bibitem{abu_hatoum_degradation_1998}
Abu~Hatoum O, Gross-Mesilaty S, Breitschopf K, Hoffman A, Gonen H, Ciechanover
  A and Bengal E 1998 {\em Molecular and Cellular Biology\/} {\bf 18}
  5670--5677

\bibitem{mcq67}
McQuarrie D~A 1967 {\em J.~of Applied Probability\/} {\bf 4} 413--478

\bibitem{gil01}
Gillespie D~T 2001 {\em J.~of Chemical Physics\/} {\bf 115} 1716--1733

\bibitem{ghaemi_stochastic_2012}
Ghaemi R and Del~Vecchio D 2012 Stochastic analysis of retroactivity in
  transcriptional networks through singular perturbation {\em American Control
  Conference (ACC)\/} pp 2731--2736

\bibitem{bokes2012multiscale}
Bokes P, King J, Wood A and Loose M 2012 {\em J. Math. Biol.\/} {\bf 65}
  493--520

\bibitem{cai2006spe}
Cai L, Friedman N and Xie X 2006 {\em Nature\/} {\bf 440} 358--62

\bibitem{friedman2006lsd}
Friedman N, Cai L and Xie X 2006 {\em Phys. Rev. Lett.\/} {\bf 97} 168302

\bibitem{sih10a}
Singh A and Hespanha J~P 2010 {\em Phil. Trans. R. Soc. A\/} {\bf 368}
  4995--5011

\bibitem{bokes2014protein}
Bokes P and Singh A {\em Submitted for publication\/}

\bibitem{bokes2013transcriptional}
Bokes P, King J, Wood A and Loose M 2013 {\em B. Math. Biol.\/} {\bf 75}
  351--371

\bibitem{Gillespie76}
Gillespie D~T 1976 {\em J.~of Computational Physics\/} {\bf 22} 403--434

\bibitem{RNC:RNC1017}
Hespanha J~P and Singh A 2005 {\em International Journal of Robust and
  Nonlinear Control\/} {\bf 15} 669--689

\bibitem{sih10}
Singh A and Hespanha J~P 2011 {\em IEEE Trans.~on Automatic~Control\/} {\bf 56}
  414--418

\bibitem{gov06}
Gomez-Uribe C~A and Verghese G~C 2007 {\em J.~of Chemical Physics\/} {\bf 126}

\bibitem{lkk09}
Lee C~H, Kim K and Kim P 2009 {\em J.~of Chemical Physics\/} {\bf 130} 134107

\bibitem{gou05}
Goutsias J 2007 {\em Biophysical Journal\/} {\bf 92} 2350--2365

\bibitem{gillespie2009}
Gillespie C~S 2009 {\em IET systems biology\/} {\bf 3} 52--58

\bibitem{Kampen01}
Kampen N~G~V 2001 {\em Stochastic Processes in Physics and Chemistry\/}
  (Amsterdam, The Netherlands: Elsevier Science)

\bibitem{singh_consequences_2012}
Singh A and Bokes P 2012 {\em Biophysical Journal\/} {\bf 103} 1087--1096

\bibitem{sos08}
Shahrezaei V, Ollivier J~F and Swain P~S 2008 {\em Molecular Systems Biology\/}
  {\bf 4} 196

\bibitem{hip11}
Hilfinger A and Paulsson J 2011 {\em Proceedings of the National Academy of
  Sciences\/} {\bf 108} 12167--12172

\bibitem{ses02}
Swain P~S, Elowitz M~B and Siggia E~D 2002 {\em Proceedings of the National
  Academy of Sciences\/} {\bf 99} 12795--12800

\bibitem{moh01}
Singh A and Soltani M 2013 {\em {PLoS} {ONE}\/} {\bf 8} e84301

\end{thebibliography}

\end{document}